\documentclass[12pt]{article}
\usepackage{axodraw}
\newcommand{\beq}{\begin{equation}}
\newcommand{\eequ}{\end{equation}}
\newcommand{\eeq}{\end{equation}}
\def\bea{\begin{eqnarray}}
\def\eea{\end{eqnarray}}
\def\as{\relax\ifmmode\alpha_s\else{$\alpha_s${ }}\fi}
\def \pt{\relax\ifmmode{p_t}\else{$p_t${ }}\fi}
\def\nn{\nonumber}
\newcommand{\noi}{\noindent}

\newif\ifdtup

\def\eqal2#1{\,\vcenter{\openup1\jot
\caja   \ialign{\strut \hfil$\displaystyle{##}$&\hfil$
\displaystyle{{}##}$\hfil &$
\displaystyle{{}##}$\hfil\crcr#1\crcr}}\,}

\begin{document}
\title{\begin{flushright}\normalsize
  \vspace{-1cm}
DESY 00-148  \\LPT Oran 00-12 \\ hep-ph/0010121
\vspace{0.5cm}
\end{flushright}
\bf The decay of the observed $J^{PC}=1^{-+}$ (1400) and $J^{PC}=1^{-+}$(1600) hybrid candidates \footnote{Work supported partially by the Abdus Salam ICTP in Trieste, Italy}}
\author{{F. Iddir$^a$\footnote{e-mail~: iddir@mail.univ-oran.dz} and A. S. Safir$^b$}\footnote{e-mail~: safir@mail.desy.de}}\par \maketitle{$^a$Laboratoire de Physique Th\'eorique, Universit\'e d'Oran Es Senia - 31100, Alg\'erie\par {$^b$DESY, Deutsches Elektronen-Synchrotron, D-22603 Hamburg, Germany}

\begin{abstract}

We study the possible interpretation of the two exotic resonances $J^{PC}= 1^{-+}$ at 1400 and 1600 MeV, claimed to be observed by BNL, decaying respectively into $\eta\pi$, $\eta'\pi$, $f_{1}\pi$ and $ \rho\pi$.
These objects are interpreted as hybrid mesons, in the quark-gluon constituent model using a chromoharmonic confining potentiel.

The quantum numbers  $J^{PC}I^{G} = 1^{-+} 1^{-}$ can be considered in a constituent model as an hybrid meson ($\,q \bar q g$). The lowest $J^{PC}= 1^{-+}$states may be built in two ways : $l_{g}$=1 (gluon-excited) corresponding to an angular momentum between the gluon and ($\,q \bar q$) system, while $l_{\,q \bar q}=1$  (quarks-excited) corresponds to an angular momentum between $q$ and $\bar q$. For the gluon-excited mode $1^{-+}$ hybrids, we find the decay dominated by the $b_{1}\pi$ channel, and by the $\rho \pi $ channel for the quark-excited mode. In our model, neither the quark-excited nor the gluon-excited  $1^{-+}$ (1400 MeV) hybrids can decay into $\eta\pi$ and $\eta'\pi$, in contradiction with experiment. Hence, the 1400 MeV resonance seems unlikely to be an hybrid state. The $1^{-+}$ (1600 MeV) gluon-excited hybrid is predicted with too  large a total decay width, to be considered as an hybrid candidate. On the contrary the quark-excited mode has a total decay width around 165 MeV, with a  $\rho \pi $ preferred decay channel, in agreement with BNL. Our conclusion is that {\it{this resonance may be considered as a hybrid meson in the quark-excited mode}}.
\end{abstract}

\newpage \pagestyle{plain}

\section{Introduction}

\hspace*{\parindent} Quantum Chromodynamics, aknowledged as the theory of strong interactions, allows that mesons containing constituent gluons ($\,q \bar q g$ hybrids) may exist. The physical existance of these "exotic" particles (beyond the quark model) is one of the objectives of experimental projects. The best clear-cut signal for an hybrid meson is to reach for $J^{PC}$ quantum numbers, not allowed in the naive quark model, such as $J^{PC}$= $0^{+-}$, $0^{--}$, $1^{-+}$, $2^{+-}$,........... From experimental efforts at IHEP \cite{Alde1}, KEK \cite{Aoyagi}, CERN \cite{Alde2} and BNL \cite{Thompson} several hybrid candidates have been identified, essentially with quantum numbers $1^{-+}$. COMPASS \cite{Moinester} program at CERN projects (through $ \gamma $-hadrons Primakoff interactions using 50-280 GeV/c hadron beams with a virtual photon target) to produce $1^{-+}$ hybrid mesons near 1.4 GeV.

In reference \cite{Barnes}, Barnes proposed a future search for exotic mesons at CEBAF and gave the hybrids which may be easily photoproduced (diffractively or by pion exchange). In this research program, the hybrids with quantum numbers $J^{PC}$= $0^{+-}$,$2^{+-}$, and $1^{-+}$ would be accessible.

In reference \cite{Achasov}, it is shown that exotic states with $I^G$($J^{PC}$)=$2^+$($2^{++}$) around 1600 MeV should be looked for (by photoproduction) at JLAB, decaying into $\rho^{\pm}$$\rho^0$, in the reactions $\gamma\ N\rightarrow \rho^{\pm}$ $\rho^0$ N and $\gamma\ N\rightarrow \rho^{\pm}$$\rho^0$$\Delta$. The aim of the $1^{-+}$ search is to study the $\eta\pi$, $\eta'\pi$ and $\rho\pi$ decay channels, to see if the E852-BNL states $1^{-+}$(1400) and $1^{-+}$(1600) are evident. In reference \cite{Blackett}, G. R.Blackett et al report on a study of the reaction $\gamma\ p\rightarrow \ p \pi^{+}\pi^{+}\pi^{-}\pi^{-}\pi^{0}$ with photon incident at 19.3 GeV, to search for resonances decaying to $b_1^{\pm} \pi^{\mp}$ (through $\gamma\  p\rightarrow \Delta^{++}\ b_1^{-}$) at SLAC. 

In reference \cite{Lie}, it is proposed to search for the hybrid $1^{-+} 1^{-}$
in the process $J/\psi \rightarrow \rho\ w \pi\pi$ at upgraded BEPC/BES, where the hybrid would be observed in the $b_{1}\pi$ channel. In reference\cite{Afanasev}, the possibility of production of an exotic $J^{PC}= 1^{-+}$ in the  $\rho^{0}\pi^{+}$ channel in the reaction $\gamma\ p\rightarrow \rho^{0} \pi^{+} \ n $ is discussed. These experimental programs would contribute significantly to the future investigation of QCD exotics, and should improve our understanding of hybrid physics.

Hybrids have been studied, using the flux-tube model\cite{Isgur}, the quark model with a constituent gluon\cite{Iddir et Olivier, Safir, Kalashnikova}, the MIT bag model\cite{Barnes et Close}, the Lattice gauge theory \cite{Michael et Mac Neike}, and QCD Sum rules\cite{Balitsky}. These models predicted that the lightest hybrid mesons will be manifested in 1.5-2 GeV mass range, a charmed hybrid meson with a mass around 4 GeV, and a bottom hybrid meson with a mass around 10 GeV.
In reference \cite{Kalashnikova}, the hybrid mesonic excitations are studied in the framework of Vacuum Background Correlator method \cite{Dosh}, where the constituent gluon is introduced starting from the perturbation theory expansion in non-perturbative QCD Vacuum \cite{Abbot}. In this method, considering the mass of quarks m(q=u,d)=0.1 GeV, m(s)=0.25 GeV, m(c)=1.5 GeV, hybrid states with possible quantum numbers $J^{PC}=0^{\pm+}$, $1^{\pm+}$, $2^{\pm+}$, $1^{\pm-}$, are predicted with masses M($\,q \bar q g$)=1.7 GeV, M($\,s \bar s g$)=2.0 GeV and M($\,c\bar c g$)=4.1 GeV. 

For exotic $J^{PC}$ hybrid mesons with quantum numbers like $0^{+-}$, $1^{-+}$, $2^{+-}$,......, considering additional gluonic excitations in a $\,q \bar q$ system, and describing the gluon field by a color flux along the links of the lattice, Lattice gauge theory predicts that $1^{-+}$ meson will be the lightest exotic hybrid, and expects a mass of 4.19 GeV for $\,c \bar c g$ , 10.18 GeV for $\,b \bar b g$, and around 1.5-1.8 GeV for lightest quarks\cite{Michael et Mac Neike}.

Using the non-relativistic flux tube model of Isgur and Paton \cite{Isgur}, 
related to the phenomenological $^3 P_0$ model \cite{Oliver}, Close et al \cite{Isgur} give predictions of the mass and the decay widths for the hybrids $1^{-+}$, $0^{-+}$, $1^{--}$ and $2^{-+}$. The model considers a $\,q \bar q$ system connected by a cylindrical bag of coloured fields: "the flux tube", and the existence of hybrid states is assumed when the flux tube is excited in the presence of the coloured $\,q \bar q$ sources. The model gives the lightest hybrid ($\,q \bar q g$) masses: 1.8-1.9 GeV for $q=u$, 2.1-2.2 GeV for $q=s$, and 4.1-4.2 GeV for $q=c$; and the gap from the lightest $_1P$ hybrid level to the first orbital excitation $_1D$ is predicted to be 0.4 GeV for ($q=u, d$) and 0.3 GeV for $q=c$. For the decay of hybrids mesons, results are given by the flux-tube model and the quark-gluon constituent model, for the allowed decay channels and the respective decay widths.

In the constituent quark-gluon model \cite{Iddir et Olivier}, we predicted that
the preferred decay channel for an hybrid meson $1^{-+}$ in the gluon-excited
mode is $b_1 \pi$, a selection rule suppressing the channels into two fundamental mesons. This rule was confirmed also by the Flux-Tube model. For the $1^{-+}$ hybrid meson in the quark-excited mode, this selection rule does not occur and the decay into two fundamental mesons is allowed; then it may decay to $\rho \pi$. Concerning the possible mixing, Lattice calculations give the result that spin-exotic hybrids cannot mix with $\,q \bar q$ states but it can mix with $\,qq\bar q\bar q$ diquonia\cite{Michael et Mac Neike}.
We deduced (using the quark-constituent model) that the mixing angle between
the $1^{--}$ $\,c \bar c g$ hybrid and the charmonium meson $\,c \bar c$ is
too small, which suppresses mixing \cite{Safir}.

Several hybrid $J^{PC}$= $1^{-+}$ candidates have been identified~: by the E852 Collaboration at BNL \cite{Adams}(in $\pi^{-} p $ interactions at 18 GeV/c) a resonant signal at $1370 \pm 40\ MeV$, and $\Gamma =385\pm \ 40 \ MeV$, decaying in $\eta \pi^{-}$, $\eta^{\prime}\pi^{-} $ and $f_{1}\pi^{-}$; by the Crystall Barrel collaboration \cite {Abele} at LEAR, the same resonance in $p\bar p $ annihilation at $1400\pm \ 20 \ MeV$with a width $\Gamma=310\pm \ 50 \ MeV$, decaying in $\eta \pi $; by the GAMS Collaboration \cite{Alde2} an exotic $1^{-+}$(1400) with $\Gamma=180\ MeV$, decaying in $\eta \pi^{0} $; by the VES Collaboration \cite{Alde1} an $1^{-+}$(1405) decaying in $\eta' \pi $ and $\eta \pi$; at BNL \cite{Lee} an $1^{-+}$ resonance at 2.0 GeV, decaying in $ f_1\pi^{-}$ and $\eta (1295)\pi^{- }$, at FNAL \cite{K.D.Blackett}, at $1740\pm \ 12 \ MeV$ and $\Gamma=136\pm \ 30 \ MeV$, decaying in $ f_{1} \pi^{-}$; by the E852 Collaboration at BNL \cite{Adams} an $1^{-+}(1600)$ exotic signal at $1593\pm \ 8\ MeV$, with $\Gamma=168\pm \ 20\ MeV$, decaying in $\rho \pi $.

In this paper, we concentrate on the two exotic $1^{-+}$ resonances, claimed to be observed by E852 Collaboration at BNL~: the $J^{PC} $= $1^{-+}$(1400) \cite{Thompson} and the $J^{PC}$= $1^{-+}$(1600), and possibly interpreted as hybrid candidates.

\section{The decay process}

\hspace*{\parindent}We use the constituent quark-gluon model \cite{Safir,Iddir et Olivier,LeYaouanc,Ishida} in which the constituent gluon moves in the framework of $\,q \bar q $ pair center of mass; and the hybrid $\,q \bar q g$ decays into two mesons through the decay of the gluon in a $\,q \bar q $ pair. Two diagrams contribute to the decay of the light hybrid, into two mesons~: 

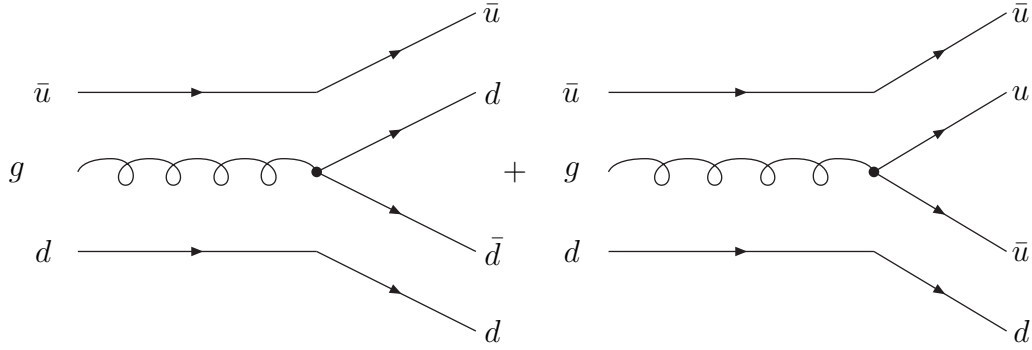
\begin{center} 
\begin{picture}(400,95)(0,0)
\ArrowLine(200,10)(300,10)
\Text(190,10)[r]{$d$}
\ArrowLine(200,70)(300,70)
\Text(190,70)[r]{$\bar u$}
\ArrowLine(300,70)(350,100)
\Text(360,100)[r]{$\bar u$}
\ArrowLine(300,40)(350,70)
\Text(360,70)[r]{$u$}
\ArrowLine(300,40)(350,10)
\Text(360,10)[r]{$\bar u$}
\ArrowLine(300,10)(350,-20)
\Text(360,-20)[r]{$d$}
\Gluon(200,40)(300,40){5}{4} \Vertex(300,40){2}
\Text(190,40)[r]{$g$}
\Text(170,40)[r]{$+$}
\ArrowLine(0,10)(90,10)
\Text(-10,10)[r]{$d$}
\ArrowLine(0,70)(90,70)
\Text(-10,70)[r]{$\bar u$}
\ArrowLine(90,70)(150,100)
\Text(160,100)[r]{$\bar u$}
\ArrowLine(90,40)(150,70)
\Text(160,70)[r]{$d$}
\ArrowLine(90,40)(150,10)
\Text(160,10)[r]{$\bar d$}
\ArrowLine(90,10)(150,-20)
\Text(160,-20)[r]{$d$}
\Gluon(0,40)(90,40){5}{4} \Vertex(90,40){2}
\Text(-20,40)[r]{$g$}
\end{picture}  
\end{center}

\vskip 5mm 
\noi with the same decay amplitude for each diagram, and a relativ sign. Then the total decay amplitude becomes the sum of the two; it can be multiplied by 2 (for example, for the  $\rho \pi $ channel), or cancelled (for  $\omega \pi $ channel). Thus, the model automatically fulfills the G-parity conservation law, in the strong interactions (the  $\omega \pi $ channel is forbidden by the G-parity).

 \section{The constituent quark-gluon model}
\label{model}
\hspace*{\parindent}
We choose the non-relativistic constituent quark model, well known to be semi-quantitatively successful in describing the hadron spectrum\cite{LeYaouanc, Iddir et Olivier, Safir}.

\noi We start from the Hamiltonian:
\beq
H = {p^2_q \over 2m_q} + {p_{\bar{q}}^2 \over 2m_{\bar{q}}} + {p_g^2 \over 2
m_g} + V\left (r_q, r_{\bar{q}}, r_g \right )
\label{1e}
\eeq
\noi where $V\left (r_q, r_{\bar{q}}, r_g \right )$ is a chromo-harmonic
potentiel easy to diagonalise.
\noi The eigenstates of the Schr\"odinger equation of (\ref{1e}), corresponding to the hybrid states are~:

\beq
\psi_{\ell_{q\bar{q}}}^{m_{q\bar{q}}}(p_{q\bar{q}}) = \left \{ {16 \pi^3
R_{q\bar{q}}^{2\ell_{q\bar{q}} + 1} \over \Gamma \left ( {3 \over 2} +
\ell_{q\bar{q}} \right )} \right \}^{1/2} \
y_{\ell_{q\bar{q}}}^{m_{q\bar{q}}}(\vec{p}_{\rho}) \ e^{- {1 \over 2} \left (
R_{q\bar{q}}^2 p_{q\bar{q}}^2 \right )} \label{11e} \eeq

\noi and

\beq
\psi_{\ell_g}^{m_g}(k) = \left \{ {16 \pi^3 R_g^{2\ell_g + 1} \over \Gamma
\left ( {3 \over 2} + \ell_g \right )} \right \}^{1/2} \ y_{\ell_g}^{m_g}(\vec{k}) e^{- {1 \over 2} R_g^2 k^2} \label{12e} \eeq\\

\noi where $y_{\ell}^m (\vec{p}) \equiv p^{\ell} Y_{\ell}^m (\theta,\Omega )$ are the harmonic polynomials.\\

\noi The classification and the decay of hybrid mesons  in a constituent model have been studied in \cite{LeYaouanc, Iddir et Olivier}~; we will use here the
notations of \cite{Iddir et Olivier}~: \\

$\ell_g$ : is the relative orbital momentum of the gluon in the $q\bar{q}$ center of mass \par
$\ell_{q\bar{q}}$ : is the relative orbital momentum between $q$ and $\bar{q}$\par 
$S_{q\bar{q}}$ : is the total quark spin \par
$J_g$ : is the total gluon angular momentum \par
$L$~: $\ell_{q\bar{q}} + J_g$. \\

\noi The parity and charge conjugation of the hybrid are given by~:
\begin{eqnarray}
\label{14e}
&&P = (-)^{\ell_{q\bar{q}} + \ell_g} \nn \\
&&C = (-)^{\ell_{q\bar{q}} + S_{q\bar{q}} + 1} \ \ \ .
\end{eqnarray}

\noi To lowest order the decay is described by the matrix element of the Hamiltonian annihilating a gluon and creating a quark pair~:

\beq
\label{19e}
H = g \sum_{ss'\lambda} \int {dp \ dk \ dp' \over \sqrt{2\omega} (2 \pi )^9} (2
\pi )^3 \ \delta (p - p' - K) \bar{u}_{ps} \ b^+_{ps} \ \gamma_{\mu} {\lambda^a
\over 2} v_{p's'} \ d^+_{p's'} \ \varphi_a \ \varepsilon_{\kappa\lambda}^{\mu} \ a_{\kappa \lambda}
\eeq

\noi in the nonrelativistic limit~:
\beq 
\bar{u}_{ps} \ \gamma_{\mu} \ v_{p's'} \ \varepsilon_{\kappa \lambda}^{\mu} =\chi_s^+ \ \sigma \ \chi_{s'} \ \varepsilon_{\kappa\lambda} \label{20e}
\eeq

\noindent where $\chi_{s'}$ is the antiquark spinor in the complex conjugate representation.
The meson states and the hybrid state are given respectively in the nonrelativistic approximation by~:

\begin{eqnarray}
|B> &=& \sum_{ss'\atop am \mu} \int {dp_q \ dp_{\bar{q}} \over (2 \pi)^6
\sqrt{3}} \ \chi_{ss'}^{\mu} (2 \pi)^3 \delta_3 \left ( p_B - p_q - p_{\bar{q}_i}\right )  \\
&& \psi \left ( {m_{\bar{q_i}} p_q - m_q p_{\bar{q}_i} \over m_q +
m_{\bar{q}_i}} \right ) <\ell_B   m_B  S_B  \mu_B | J_B   M_B> b^+_{p_q
sa} \ d^+_{p_{\bar{q}_i}s'\bar{a}}|0> \quad . \nn \label{21e}
\end{eqnarray}

\noi with $\int dp/(2 \pi)^3 |\psi_B (p)|^2 = 1$, and the same for the $|C>$.

Note here that $q$ and $\bar{q}$ can be different in flavour and we specify the creating quark pair by $q_i$ $\bar{q}_i$

\begin{eqnarray}
\label{22e}
|A> &=& \sum_{ss'\atop am\mu} \int {dp_q \ dp_{\bar{q}} dk \over (2 \pi)^3} \
\chi_{ss'}^{\mu} (2 \pi )^3 \ \delta_3 (k + p_q +
p_{\bar{q}}) {\lambda^{c_g}_{c_q \ c_{\bar{q}}} \over 4}\\
&&\psi \left ( {m_{\bar{q}} p_q - m_q p_{\bar{q}} \over m_q +
m_{\bar{q}}} \right ) \ \psi \left ( {(m_q + m_{\bar{q}})k - m_g (p_q +
p_{\bar{q}}) \over m_q + m_{\bar{q}} + m_g} \right ) \nn \\
&&<\ell_g  m_g  1  \mu_g |J_g M_g> <\ell_{q\bar{q}} m_{q \bar{q}} J_g M_g
|L {m'}> < L {m'} \ S_{q \bar{q}} \ \mu_{q \bar{q}}|JM > \nn \\
&&b^+_{p_qsa} \ d^+_{p_{\bar{q}}s'\bar{a}} \ a^+_{\kappa \lambda}|0>  \nn
\end{eqnarray}

\noi where $c_q$, $c_{\bar{q}}$ and $c_g$ are the color charge of the quark, antiquark and gluon with $c_q = 1, \cdots 3$~; $c_{\bar{q}} = 1, \cdots 3$~; $c_g= 1, \cdots 8$.

Using (\ref{19e}) one gets in a straighforward manner the matrix
element $<BC|H|A>$ between an hybrid state $A$ and two mesons $B$ and $C$~:
\beq
<BC|H|A> = gf (A, B, C) (2 \pi)^3 \ \delta_3(p_A - p_B - p_C)
\label{23e}
\eeq

\noi where $f(A,B,C)$ representing the decay amplitude by~:

\begin{eqnarray}
f(A, B, C) &=& \sum_{m_{q\bar{q}}, m_g, m_B, m_C \atop \mu_{q\bar{q}}, \mu_g,\mu_B, \mu_C} \ \Omega \phi \ X(\mu_{q \bar{q}}, \mu_g ; \mu_B , \mu_C ) \\&&I \left ( m_{q\bar{q}}, m_g ; m_B, m_C, m \right ) <\ell_g m_g 1 \mu_g|J_g M_g> \nn \\&&<\ell_{q\bar{q}} m_{q\bar{q}} J_g  M_g |L {m'}> <L {m'} S_{q\bar{q}}\mu_{q\bar{q}} |J M> \nn \\&&< \ell_B   m_B  S_B  \mu_B |J_B M_B> < \ell_C  m_C  S_C  \mu_C |J_C M_C>
\nn
\label{24e} \end{eqnarray}

\noi where $\Omega$, $X$, $I$ and $\phi$ are the color, spin, spatial and flavour overlaps. $\Omega$ is given by~:

\beq
\Omega = {1 \over 24} \sum_a {\rm Tr} \ (\lambda^a)^2 = {2 \over 3}
\label{25e}
\eeq

\noi from :
\beq
\chi_{\mu_1}^+ \ \vec\sigma \ \ \chi_{\mu_2} \ \vec\varepsilon^{\lambda}= \sqrt{3} < {1 \over 2} \mu_1 {1 \over 2} \mu_2 |\ell \lambda >\label{26e}
\eeq

\noi we obtain the spin overlap~:

\begin{eqnarray}
X(\mu_{q\bar{q}}, \mu_g, \mu_B, \mu_C) &=& \sum_s \sqrt{2} \left [\begin{array}{lll} 1/2 &1/2 &S_B \cr 1/2 &1/2 &S_C \cr S_{q \bar{q}} &1 &S
\cr \end{array} \right ] <S_{q\bar{q}} \ \mu_{q\bar{q}} 1 \mu_g |S \ \mu_{q
\bar{q}} + \mu_g> \nn \\&&<S_B \ \mu_B \ S_C \ \mu_c |S \ \mu_B + \mu_C >
\label{27e}
\end{eqnarray}

\noi where :

\beq
\left [ \begin{array}{lll} 1/2 &1/2 &S_B \cr 1/2 &1/2 &S_C \\ S_{q
\bar{q}} &1 &S \end{array}\right ] = \sqrt{(2 S_B + 1) (2S_C + 1) 3(2S_{q
\bar{q}} + 1)} \left \{ \begin{array}{lll} 1/2 &1/2 &S_B \cr 1/2 &1/2 &S_C
\cr S_{q\bar{q}} &1 &S \end{array} \right \}
\label{28e}
\eeq

\noi the term between bracketts in  eq.(\ref{28e}) is called $9j$ and corresponds to the coupling of the four spins~: $S_B, S_C, S_{q\bar{q}}$ and $S_g=1$.\\

\noi the spatial overlap is given by~:

\begin{eqnarray}
\label{29e}
&&I \left ( m_{q \bar{q}} , m_g, m_B, m_C, m \right ) = \int \int {dp \ dk \over
\sqrt{2\omega} (2 \pi)^6} \psi_{\ell_{q\bar{q}}}^{m_{q\bar{q}}} (p_B - p) \
\psi_{\ell_g}^{m_g} (k) \\
&&\psi_{\ell_B}^{m_B^*} \left ( {p_B m_{\bar{q}_i} \over m_q + m_{\bar{q}_i}} -
p - {k \over 2} \right ) \psi_{\ell_C}^{m_C^*} \left ( -{m_{q_i} p_B \over
m_{q_i} + m_{\bar{q}}} + p - {k \over 2} \right ) d\Omega_B \nn
Y_{\ell}^{m^*}(\Omega_B) \end{eqnarray}

\noi where $\ell$, $m$ label the orbital momentum between the two final
mesons.\\

Finally,

\beq
\phi = \left [ \begin{array}{lll} i_1 &i_3 &I_B \cr i_2 &i_4&I_C \cr I_A &0 &I_A \cr \end{array}\right ] \eta \varepsilon  \label{30e}
\eeq

\noi where the $I$'s ($i$'s) label the hadron (quark) isospins, $\eta = 1$ if
the gluon goes into strange quarks and $\eta = \sqrt{2}$ if it goes into
non strange ones. $\varepsilon$ is the number of diagrams con\-tri\-bu\-ting to the decay. Indeed one can check that since $P$ and $C$ are conserved, two diagrams
contribute with the same sign and magnitude for allowed decays while they
cancel for forbidden ones. In the case of two identical final particles,
$\varepsilon = \sqrt{2}$. The partial width is then given by~:

\beq
\Gamma (A \to BC ) = 4 \ \alpha_s \ |f(A, B, C)|^2 {P_B \ E_B \ E_C \over M_A}
\label{31e}
\eeq

\noi with~:
\begin{eqnarray}
&&P_B^2 = { \left [ M^2 - (m_B + m_C)^2 \right ] \left [ M^2 - (m_B - m_C)^2
\right ] \over 4M^2} \qquad ; \quad (M \equiv M_A) \nn \\&&E_B = \sqrt{P_B^2 + m_B^2} \qquad , \quad E_C = \sqrt{P_B^2 + m^2_C} \quad .
\label{32e}
\end{eqnarray}

\section{The $1^{-+}$ hybrid and the selection rules}

\subsection{Classification of the $1^{-+}$ hybrid states}

\hspace*{\parindent}

Let us consider the lightest $J^{PC}=1^{-+}$ hybrid mesons, using the quark model with a constituent gluon \cite{LeYaouanc, Iddir et Olivier, Safir, Ishida}, as developped in section 3. Eq.(\ref{14e}) implies $\ell_{q\bar{q}}$+$\ell_g$ odd and $\ell_{q\bar{q}}$ + $S_{q\bar{q}}$ odd. The lightest such states are $S_{q\bar{q}}$=0, $\ell_{q\bar{q}}$=1 and $\ell_g$=0, which we shall refer to as the quark-excited hybrid ($\ell_{q\bar{q}}$=1), and $S_{q\bar{q}}$=1, $\ell_{q\bar{q}}$=0 and $\ell_g$=1, which we shall refer to as the gluon-excited hybrid ($\ell_g$=1). In the case of the gluon-excited one (respectively the quark-excited one) we obtain from the Clebsch Gordan of the Eq.(\ref{22e})~: L=$J_g$=0,1,2 (respectively L=J=$J_g$=1)

\begin{table}[htb]
\begin{center}
\begin{tabular}{|c|c|c|c|c|c|c|c|}
\hline
& & & & & & &\\
$P$ & $C$ & $\ell_{q\bar{q}}$ & $\ell_g$ &  $J_g$ & $S_{q\bar{q}}$ & $L$ & $J$\\
& & & & & & &\\
\hline
$-$ & $+$ & 0 & 1 & 0 & 1 & 0 & 1 \\
$-$ & $+$ & 0 & 1 & 1 & 1 & 1 & 1 \\
$-$ & $+$ & 0 & 1 & 2 & 1 & 2 & 1 \\
$-$ & $+$ & 1 & 0 & 1 & 0 & 1 & 1 \\
\hline
\end{tabular}
\label{tab1}\vskip 5mm
\caption{\leftskip 1pc \rightskip 1pc
\baselineskip=10pt plus 2pt minus 0pt Lowest $J^{PC}=1^{-+}$ $\,q \bar q g$ hybrid mesons and their quantum numbers}
\end{center}
\end{table}

Restricting ourselves to the lightest states ($\ell_{q\bar{q}}$+$\ell_g$=1), we have listed in Table 1 three hybrid mesons with an orbital excitation between the gluon and the ${q\bar{q}}$ system ($\ell_g$=1), and one with an orbital excitation between the q and the ${\bar{q}}$ ($\ell_{q\bar{q}}$=1).

\subsection{The $1^{-+}$ selection rules for the decay}

\hspace*{\parindent}
Let us compute the integral  (\ref{29e}), respectively for $\ell_B$=0, $\ell_C$=0 and $\ell_B$=1, $\ell_C$=0, assuming $R_B=R_C$, we obtain~:

\begin{eqnarray}
\label{37e}
&&I(m_{q \bar{q}}, 0; 0, 0, m) = 2^4 \sqrt{{\pi \over 3 \omega}} {R_{q\bar{q}}^{3/2+\ell_{q\bar{q}}} \ R_g^{3/2+\ell_g} \ R_B^5 \over \left (R_g^2 +R_B^2/2 \right )^{3/2}\left ( R_{q \bar{q}}^2 + 2R_B^2 \right )^{5/2}}
{2m_q \over m_q + mq_i} \ P_B \  \\
&& \exp \left\{- {P_B^2 \over 2}
\left[R_{q \bar{q}}^2 + {2 m^2_{\bar{q}_i}R_B^2 \over (m_q +m_{\bar{q}_i})^2} - { \left [ 2m_{q_i} \ R_B^2 + \left ( m_q + m_{\bar{q}_i} \right )R_{q\bar{q}}^2 \right ]^2 \over \left ( R^2_{q \bar{q}} + 2R_B^2 \right )\left( m_q + m_{\bar{q}_i} \right )^2} \right]\right\} \delta_{\ell_{g,0}}\delta_{\ell_{q \bar{q}}, \ell} \delta_{m_{q \bar{q}}, m}\nn
\end{eqnarray}

\noi and~:
\begin{eqnarray}
\label{36e}
&&I(0, m_g; m_B, 0, 0) = - \sqrt{{\pi \over 2 \omega}} {R_{q\bar{q}}^{3/2} \ R_g^{5/2} \ R_B^4 \over \left (R_g^2/2 +R_B^2/4 \right )^{5/2}\left ( R_{q \bar{q}}^2/2 + R_B^2 \right )^{3/2}}\\
&& \exp \left\{- {P_B^2 \over2}
\left[ R_{q \bar{q}}^2 + {2 m^2_{\bar{q}_i}R_B^2 \over (m_q +m_{\bar{q}_i})^2}
- { \left [ 2m_{q_i} \ R_B^2 + \left ( m_q + m_{\bar{q}_i} \right )R_{q\bar{q}}^2 \right ]^2 \over \left ( R^2_{q \bar{q}} + 2R_B^2 \right )\left ( m_q + m_{\bar{q}_i} \right )^2} \right]\right\} \delta_{\ell_{g}, 1} \delta_{\ell_{q \bar{q}}, 0} \delta_{m_{g}, m_{B}} \delta_{m_{q \bar{q}}, 0}\nn \end{eqnarray}

The first result eq.(\ref{37e}), corresponds to the quark-excited mode ($\ell_{q\bar{q}}=1$, $\ell_g=0$). In this mode, the hybrid is not allowed to decay into two mesons ($l_B=1$ + $l_C=0$) like  ${b_1 \pi^-}$, ${f_1\pi^-}$,...
Such decay could occur for waves of even parity (S,D) and the term $\delta_{\ell_{{q\bar{q}},l}}$ in eq.(\ref{37e}) implies a P-wave for the process. Then the quark-excited hybrid is allowed to decay only into two ground state mesons, like ${\rho \pi}$, ${\rho \omega}$ ,....... This is the selection rule for the quark-excited hybrids decay \cite{Safir}.

The $\delta_{\ell_{g,1}}$ term in eq.(\ref{36e}) tells us that a gluon-excited hybrid is not allowed to decay into two ground state mesons like $\eta\pi$, $\eta'\pi$,$\rho\pi$,..., but decays into two mesons respectively (1S) and (1P) states (like  ${b_1 \pi^-}$, ${f_1\pi^-}$,..). This is the well known selection rule for the gluon-excited hybrid decays \cite{Iddir et Olivier}.

\section{The decay widths}
\hspace*{\parindent}
We now consider the states in Table 1. The $J^{PC}= 1^{-+}$ hybrid has $I^{G}= 1^{-}$ (Isospin and G-parity).
Using the conservation of angular momentum, parity, isospin and G-parity, we give the predicted decay channels for the two modes (quarks-excited and gluon-excited).

\subsection{Decay widths of the $1^{-+}$ quark-excited hybrids}
\hspace*{\parindent}
In the case with ($\ell_{q\bar{q}}=1$, $l_g=0$), the hybrid meson decays into two (L=0) mesons, and the allowed channels are:  $\rho \omega $, $\rho(1450) \pi$, $\rho \pi $ and ${K^*\ K}$. The decay is dominated by $\rho \pi $, both for the $1^{-+}$ (1400) and the (1600) MeV.
Table 2 shows the 1.4 GeV and 1.6 GeV, $1^{-+}$ quark-excited hybrids decay widths. 

In our calculations, we take~: $R_B^2$ = $R_C^2$ = $R_g^2$ =$ 6 \ {\rm GeV}^{-2}$ , \ $R_{q\bar{q}}^2 = 12 \ {\rm GeV}^{-2}$,
\noindent  $m_s$ = 0.5  \ {\rm  GeV} ,\ $m_u$ = $m_d$ = 0.35 \ {\rm GeV}\ and \
$m_g$ = 0.8 \ {\rm GeV}.  

\vskip 5mm
\begin{table}[htb]
\begin{center}
\begin{tabular}{|c|c|c|}
\hline 
$ $ & $1^{-+}(1400)$ & $1^{-+}(1600)$ \\ \hline 
$\Gamma_{\rho\omega}$  & &4 \\ \hline 
$\Gamma_{\rho\pi}$  &72 & 142\\ \hline
$\Gamma_{K^*\ K}$ &0.3 &19\\ \hline 
$\Gamma_{tot}$ &72.3 &165 \\ \hline
\end{tabular}
\end{center}

\vskip 0 cm
\caption{\leftskip 1pc \rightskip 1pc
\baselineskip=10pt plus 2pt minus 0pt
{\sl Predicted widths in }$\alpha_s$ MeV for the decay of a quark-excited ($\ell_{q\bar{q}}$=1) $1^{-+}$ hybrid meson of mass 1.4 and 1.6 GeV }
\label{tab2}\vskip 5mm
\end{table}
\vskip 5 truemm

\subsection{Decay widths of the $1^{-+}$ gluon-excited hybrids}
\hspace*{\parindent}
In the case where ($l_g=1$, $\ell_{q\bar{q}}=0$) (gluon-excited mode), due to the selection rule suppressing the decay into two fundamental mesons, the allowed decay channels are~: $ b_1^0 \pi^-$and $ b_1^- \pi^0$ for the $1^{-+}$(1400) and $ b_1^0 \pi^-$, $ b_1^- \pi^0$, $f_1(1285) \pi^-$and $f_1(1420) \pi^-$for the $1^{-+}$(1600).
Then, the decay occurs into the channel $ b_1 \pi$ for the $1^{-+}$(1400), and is dominated by  $ b_1 \pi$ for the $1^{-+}$(1600).
The quark-gluon constituent model predictions for the decay of a 1.4 GeV and 1.6 GeV  $1^{-+}$ gluon-excited hybrids are given respectively in Table 3 and Table 4. We take in our calculations the same masses and parameters as for the decay of the quark-excited hybrid, in the preceding section.

\vskip 5mm
\begin{table}[htb]
\begin{center}
\begin{tabular}{|c|c|c|c|}
\hline
$L$ & $0$ & $1$ & $2$ \\ \hline
$\Gamma_{b_1^0\pi^-}$ & 44 & 131 & 220 \\ \hline
$\Gamma_{b_1^-\pi^0}$ & 47 & 141 & 235 \\ \hline
$\Gamma_{tot}(1.4)$ & 91 & 272 & 455 \\ \hline
\end{tabular}
\end{center}

\vskip 0 cm \caption{\leftskip 1pc \rightskip 1pc
\baselineskip=10pt plus 2pt minus 0pt{\sl Predicted widths in $\alpha_s$ MeV for the decay of a gluon-excited ($l_g=1$) $1^{-+}$ hybrid meson of mass $(1.4)$ GeV} }
\label{tab3}\vskip 5mm
\end{table}
\vskip 5 truemm

\vskip 5mm
\begin{table}[htb]
\begin{center}
\begin{tabular}{|c|c|c|c|}
\hline
$L$ & $0$ & $1$ & $2$ \\ \hline
$\Gamma_{b_1^0\pi^-}$ & 230 & 688 & 1146 \\ \hline
$\Gamma_{b_1^-\pi^0}$ & 230 & 690 & 1150 \\ \hline
$\Gamma_{f_1(1285)\pi^-}$ & 186 & 140 & 234 \\ \hline
$\Gamma_{f_1(1420)\pi^-}$ & 62 & 46 & 76 \\ \hline
$\Gamma_{tot}(1.6)$ & 708 & 1564 & 2606 \\ \hline
\end{tabular}
\end{center}

\vskip 0 cm
\caption{\leftskip 1pc \rightskip 1pc
\baselineskip=10pt plus 2pt minus 0pt
{\sl Predicted widths in $\alpha_s$ MeV for the decay of a gluon-excited ($l_g=1$) $1^{-+}$ hybrid meson of mass $(1.6)$ GeV} }
\label{tab4}\vskip 5mm
\end{table}
\vskip 5 truemm

\vskip 5mm 

\section{Discussion and Conclusion}
\hspace*{\parindent} We have considered the $J^{PC}I^{G} = 1^{-+} 1^{-}$ hybrid ($\,q \bar q g$) candidates with masses 1400 MeV and 1600 MeV. Such objects are claimed to be observed at BNL. We used the quark-gluon constituent model, implemented with chromoharmonic confining potentiel. Considering the quantum numbers of these candidates, and restricting ourselves to the lightest states ($l_g+l_{q\bar q}$=1), we have found three hybrids with an orbital excitation between the gluon and the $q\bar q$ system (this is the gluon-excited mode~: $l_g=1$), and one with an orbital excitation between  the $q$ and the $\bar q$ (this is the quark-excited mode~: $l_{q\bar q}$=1).

In our model, the decay of the hybrid occurs through two diagrams, in which the gluon decays to a $\,q \bar q$ pair. The model automatically fulfills the G-parity conservation law, in the strong interactions (the  $\omega \pi $ channel is forbidden by the G-parity, and the decay amplitude for this process gives zero).

Considering the conservation laws of angular momentum, parity, charge conjugation, isospin, G-parity, and the selection rules for the decay, we find the following results for the two resonances~:

- the decay of the quark-excited hybrid is allowed only into two ground-state mesons ($l_B=0,l_C$=0) and is dominated by  $\rho \pi$ for both $1^{-+}$ 1400 MeV and 1600 MeV;

-the decay of the gluon-excited hybrid occurs only into two mesons, whose one fundamental and one orbitally excited  ($l_B=1,l_C$=0); then $(b_1\pi)$ dominates the decay.

We find a total decay width around 70 MeV for the quark-excited  $1^{-+}$ (1400 MeV), and 90 MeV for the gluon-excited one. For the  $1^{-+}$ (1600 MeV), the quark-excited mode has a total decay width of 165 MeV, and the gluon-excited mode  around 700 MeV for L=0 and even worst for $L\ge 1$.
Then for light hybrids (of mass 1400 MeV or 1600 MeV), our model predicts total decay widths small enough as compared to the level spacing, to generate observable resonance, except for the gluon-excited hybrid 1600 MeV.

The BNL experiment observed the $1^{-+}$ (1400) resonance into $\eta \pi^-$, $\eta^{\prime}\pi^- $ and $f_1 \pi$, with a total width of $\sim$ 400 MeV. In our model, the hybrid candidate cannot decay in $\eta \pi^-$ and $\eta^{\prime}\pi^- $, in either of the two cases~: the gluon-excited mode due to the selection rule which suppresses the decay into two fundamental mesons, and the quark-excited mode due to the non-conservation of the total spin in the final state. The $f_1 \pi$ channel is not allowed in our model due to the limited phase-space since we assume the narrow resonance approximation. Clearly a correct consideration of resonance widths would allow this channel.

The $1^{-+}$ (1600) resonance has been observed into $\rho \pi $, with a width $\Gamma= 168 \pm \ 20 \ MeV$. We found in our model this channel with $\Gamma= 142 \ MeV$, and also ${K^*\ K}$ ( $\Gamma=19 \ MeV$) and $\rho \omega$ (4 MeV). This result means that {\it{this resonance may be considered as an hybrid meson in the quark-excited mode}}. Indeed, we predict for the quark-excited mode the observed decay channel, with a reasonable width in good agreement with BNL. 

It should be noted that a more realistic potential might induce a large mixing between the two classes of hybrids, which will be totally dominated by the large hybrid class width. In fact this could explain our total predicted width of $\sim$ 100 MeV for our 1400 MeV hybrid candidate which is in order of magnitude with the flux-tube predictions ($\sim$ 110 MeV) \cite{Donnachie et Page}.
In such a realistic potential model, our  1600 MeV total predicted width will be totally dominated by the gluon-excited mode leading to a very large total width making him a bad hybrid candidate. Our prediction for this later is totally different from ref.\cite{Donnachie et Page}, where the authors give the total decay widths of the orbitally excited hybrid 1600 MeV ($\sim$ 150 MeV) to P+S wave  states. It is unreasonable to assume some similarity between the gluon-excited hybrid and the orbitally excited hybrid in the flux-tube model since they do decay to the same channels, but the predicted widths are quite differents.
However  they do not give any result concerning their P-wave modes which is the quark-excited mode in our model (see Table \ref{tab4}), and this is due exclusively to their selection rule for the lowest orbitally excited hybrid.

It is not casually that in a more realistic potential the two models  predict the same order of magnitude for the 1400 MeV hybrid candidate and this is just an indication for a model independant result for this candidate.

Unhappily, it becomes difficult to understand  the $\sim$ 400 MeV width of the E852 Coll.\cite{Thompson} as an evidence for an exotic resonance at 1400 MeV, which is totally forbidden in our model.
Ref.\cite{Donnachie et Page} rejects the hypothesis that the 1400 and 1600 MeV states are both hybrid mesons, because of the experimental width which is larger in the 1400 MeV state than the 1600 MeV state which should be opposite.  They interpret the experimental observed peak in the $\eta \pi$ channel, which is suppressed by symmetrization selection rules, as a sizable final state interactions and they suggest that the E852  $\eta \pi$ peak is due to the interference of a Deck-type background with a hybrid resonance of higher mass, for which the $\hat{\rho}$ at 1.6 GeV is an obvious candidate.                         

Finally to check the existence of the hybrid mesons, namely the $1^{-+}$ ones, and have more informations about the physics of these objects, it will be important to have (from the experiments) the observed branching ratio in $\eta \pi$ and $\eta^{\prime}\pi$. Furthermore, it will be necessary to check the channel $b_1\pi$, in order to verify the selection rule concerning the gluon-excited hybrid, which suppresses the decay into two fundamental mesons, and allows the decay into one fundamental meson and one orbitally excited meson.



\vskip 5mm 

\section*{Acknowledgements}

\hspace*{\parindent}We are grateful to O. P\`{e}ne for several stimulating and illuminating discussions. A. S. S. would like to thank the German Academic Exchange Service (DAAD) for financial support, and the Abdus Salam International Centre for Theoretical Physics, in Trieste (Italy), for their Fellowship to visit the Centre, where a big part of this work was done and DESY, where this work was completed.
 
F. I. acknowledges the Laboratoire de Physique Th\'eorique of the Paris-Sud (Orsay-France) University for its hospitality, and the Abdus Salam ICTP, where this  work was completed (under the Associateship Scheme).

\end{document}